\documentclass{ptapap}

\def\d{\,d$^{-1}$}

\author{Thomas Kallinger}[IFA]
\author{Werner W. Weiss}[IFA]
\author{the BRITE Team}[]
\affil[IFA]{Institute for Astrophysics (IfA), University of Vienna, T\"urkenschanzstrasse 17, 1180 Vienna, Austria}

\title{Testing the photometric stability of BRITE-Constellation}

\begin{document}

\maketitle

\begin{abstract}
To characterise the long-term stability and true photometric sensitivity of a space-based instrument is a difficult task and can be done best with independent measurements of a relatively quiet star. A rare occasion for such a test are the complementary observations of some bright Plejades stars with the \textit{Kepler}/K2 telescope and the BRITE-Austria (BAb) and UniBRITE (UBr) satellites. While most of them show a complex oscillatory behaviour, the frequency spectrum of the bright B-type star Atlas is relatively simple. From the 71 days-long K2 observations we extract the three dominant frequencies and show that the residuals have a noise level of less than 10ppm. While the BAb observations are not sensitive enough, we find the same periodicities in the 167 days-long UBr data set, which shows no additional significant signal down to a noise level of about 100ppm. This impressively demonstrates the stability of the BRITE instruments in the low-frequency regime and their capability to detect sub-mmag variability, even for stars close to the faint end of the nominal dynamic range.
\end{abstract}

\section{Introduction}
The BRITE-Constellation\footnote{BRITE-Constellation consists of five operational nano-satellites: BRITE-Austria (BAb) and BRITE-Lem (BLb) equiped with a ``blue'' filter and UniBRITE (UBr), BRITE-Heweliusz (BHr), and BRITE-Toronto (BTr) equiped with a ``red'' filter.} satellites \citep{weiss2014} are now operational for more than four years and have since delivered millimagnitude-precision time-series photometry in two passbands for a large fraction of all stars in the sky brighter than 4th magnitude. Like any other space-based photometer, they do not switch between target star and a constant comparison star so that it is difficult to characterise the long-term stability and true photometric sensitivity of the instruments. 

A rare occasion for testing this are the (almost) simultaneous observations of several bright stars in the Plejades with BAb/UBr and with the \textit{Kepler}/K2 space telescope \citep{bor10,Howell2014}. Among those, the 3.6\,mag bright star Atlas (27 Tau, HD23850) shows a light curve with mmag variability on timescales on the order of half a day to two days \citep{White2017}. The structure of the intrinsic signal is, however, quite simple and can easily be pre-whitened from the data so that Atlas is well-suited for a comparison of the photometric performance of the different instruments. The origin of the variability is not yet understood since Atlas is a visual binary system consisting of two stars of spectral type B8 with a relatively long orbital period of $\sim$271\,d \citep{Zwahlen2004}. While the brighter component is a giant for which no oscillations are expected, the $\sim$1.7\,mag fainter component is a main-sequence star that is located right in the middle of the slowly pulsating B-star (SPB) instability region in the HR-diagram. It seems therefore plausible that the detected variability is due to the high-order gravity modes of a SPB star \citep{Waelkens1991}.

\begin{figure}[h]
\centering
\includegraphics[width=1.0\textwidth]{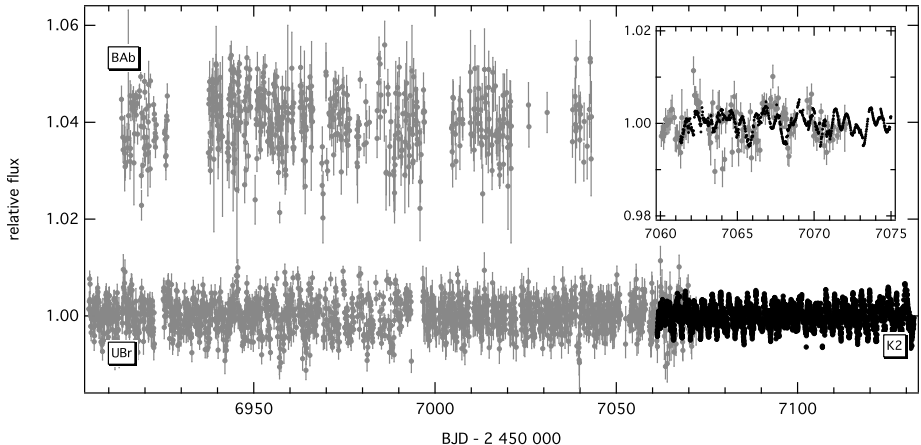}
\caption{Final light curve of Atlas as obtained with BAb (arbitarily shifted in y by 0.04), UBr, and K2. While the BRITE data are binned to one data point per satellite orbit, the full K2 time series is shown. The insert zooms onto the about 11 days with simultaneous UBr and K2 observations.}
\label{fig:LC}
\end{figure}

\section{K2 observations}
The nominal \textit{Kepler} mission ended with the loss to maintain stable pointing in its original field due to a failure of two reaction wheels. However, by aligning the telescope along its orbital plane, the now K2 named mission can control stable pointing in fields along the ecliptic for up to $\sim$80\,d by using the remaining two reaction wheels and periodic firings of the thrusters.

The \textit{Kepler} CCDs saturate for stars with $K_p<11-12$\,mag, with excess flux bleeding along CCD columns. To recover the light curves of bright stars therefore requires special methods. One of them was developed by \cite{White2017}, who use the flux recorded by non-saturated pixels in the halo surrounding a bright star. They extracted the light curves of the seven brightest stars in the Plejades (including Atlas) from the K2 observations, which were obtained during Campaign 4 between Feb. -- Apr. 2015. The about 71 days-long continuous time series of Atlas consists of $\sim$2390 measurements with a cadence of 29.4\,min and is shown in Fig.\,\ref{fig:LC}. 

\begin{figure}[h]
\centering
\includegraphics[width=1.0\textwidth]{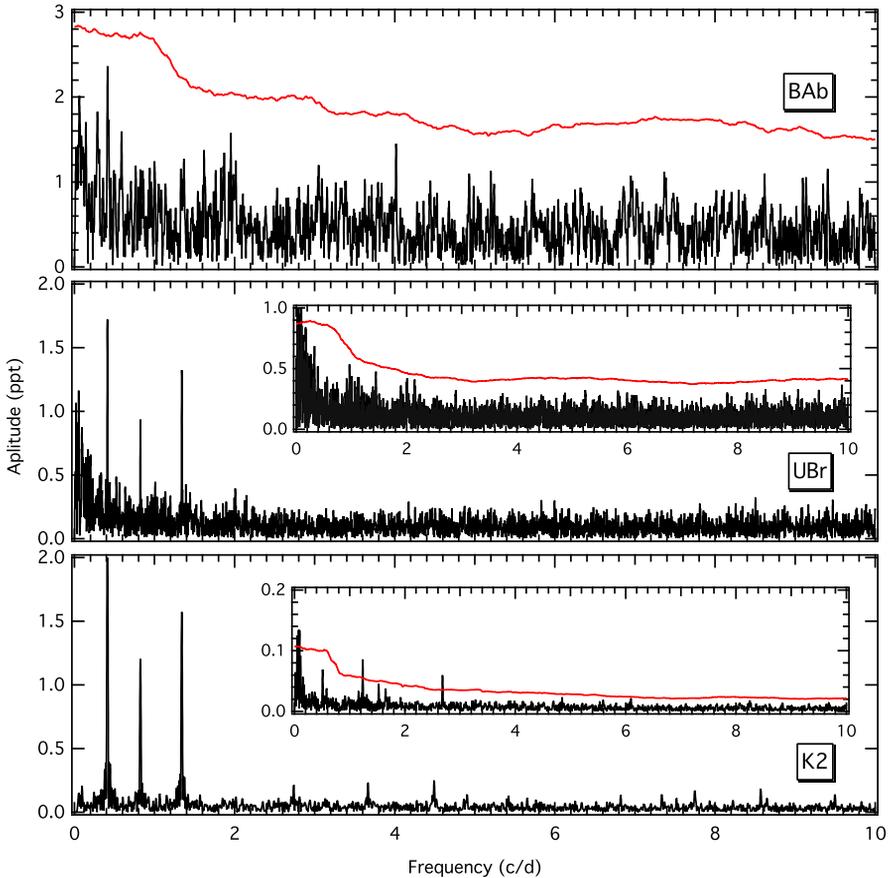}
\caption{Amplitude spectrum of the BAb (top), UBr (middle), and K2 (bottom) observations of Atlas. The inserts show the residual spectra after pre-withening of the three dominant frequencies. The red lines correspond to the local detection limit (SNR$>4$).}
\label{fig:fou}
\end{figure}

The Fourier amplitude spectrum of the K2 data set is illustrated in the bottom panel of Fig.\,\ref{fig:fou} and is dominated by three peaks in the low-frequency regime. After pre-whitening we find a few more peaks with amplitudes below 100\,ppm and a noise level of about 6\,ppm. A full list of significant frequencies may be found in \cite{White2017}. For our purpose we can ignore these low-amplitude peaks so that it is clear that Atlas does not show -- apart from the three dominant frequencies -- intrinsic signal with amplitudes larger than about 100\,ppm. The star is therefore in fact a good target for testing the photometric performance of BAb and UBr.

\section{BRITE observations}
Between Sep. 2014 and Feb. 2015, Atlas was observed almost continuously with UBr for typically 17\% of the 101\,min satellite orbit with a cadence of about 15\,s resulting in a total of more than 165\,000 individual measurements covering about 168 days. BAb used a different setup with more gaps leading to total of about 25\,500 measurements obtained during a total of about 130\,d. Details about the detector and data acquisition of BRITE-Constellation are described by \cite{pablo2016}.

The raw data were reduced as describe by \cite{popowicz2016} and remaining instrumental signals were corrected following the recipe of  \cite{Kallinger2017}. The final light curves are shown in Fig.\,\ref{fig:LC}, where we bin the data per satellite orbit for better visibility. The bins are  based on typically 38 and 20 points of the UBr and BAb time series, respectively, and the mean brightness scatters on average by about 1.9\,ppt and 3.1\,ppt.

In Fig.\,\ref{fig:fou} we show the Fourier amplitude spectra of the two data sets in the range of 0--10\d . Above about 12\d\ the spectrum is dominated by alias signal originating from the gaps in the time series regularly modulated with the 101\,min orbit of the satellite (14.2\d).

For the BAb data we find no peak with an amplitude exceeding the formal significance criterion (red lines in Fig.\,\ref{fig:fou}) of a signal-to-noise ratio (SNR) $>4$ \citep[e.g.,][]{kuschnig1997}. The largest peak in the spectrum ($\sim$0.4\d ; SNR$\simeq$3.5) , however, very likely corresponds to the dominant peak in the intrinsic spectrum (see next Sec.). The spectrum has a noise level of about 0.4-0.7\,ppt so that the detection limit is on the order of 1.6-2.8\,ppt, depending on the frequency.

The quality of the UBr data is much better, as can be seen from the insert in Fig.\,\ref{fig:LC} (where we show the about 11\,d for which simultaneous UBr and K2 observations are available) and the middle panel of Fig.\,\ref{fig:fou}. We can identify three peaks with a SNR$>4$ that clearly correspond to the three dominant frequencies of Atlas. After pre-whitening them there is only some long-period ($<$0.1\d ) signal left, which is likely due to residual instrumental long-term instabilities. The noise level in the residual spectrum is on the order of 100\,ppm leading to a detection limit of 0.4-0.5\,ppt, depending on the frequency. We furthermore note that the three extracted frequencies agree in all parameters within the uncertainties with those reported by \cite{White2017} based on the K2 data.

\section{Conclussions}
Thanks to the high-precision K2 photometry of the bright Plejades star Atlas we can for the first time directly test the photometric sensitivity of two BRITE satellites. We find that both satellites perform within the mission requirements. While BAb might not the be the right instrument for sub-millimag variability, UBr can in fact detect intrinsic low-frequency signals with amplitudes as low as 0.5\,ppt, even for stars close to the faint end of the nominal dynamic range. We also find no indications for any instrumental artefacts in the investigated frequency range.

Apart from that, the combination of BRITE and K2 observations has also some scientific value. By combining the two time series we can extend the total observing time to about 228\,d. We then find the three dominant peaks to be pairs of close frequencies (which are not resolvable in the individual data sets) typical for rotational splittings in SPB stars. For more details along with the analysis of other Plejades stars we refer to Kallinger et al. (in prep.).

\acknowledgements{The authors are grateful for funding via the Austrian Space Application Programme (ASAP) of the Austrian Research Promotion Agency (FFG) and BMVIT. The paper is based on data collected by the BRITE Constellation satellite mission, designed, built, launched, operated and supported by the Austrian Research Promotion Agency (FFG), the University of Vienna, the Technical University of Graz, the Canadian Space Agency (CSA), the University of Toronto Institute for Aerospace Studies (UTIAS), the Foundation for Polish Science \& Technology (FNiTP MNiSW), and National Science Centre (NCN).
}

\bibliographystyle{ptapap}
\bibliography{PhotStability}

\end{document}